\input{aipcheck}

\documentclass[
    ,final            
  ]
  {aipproc}

\layoutstyle{6x9}

\begin{document}

\title{The LHCb Upgrade}

\classification{}
\keywords      {flavour physics, instrumentation, detector upgrade}

\author{Lars Eklund, on behalf of the LHCb Collaboration}{
  address={School of Physics \& Astronomy, Kelvin Building, University
  of Glasgow, Glasgow G12 8QQ, UK}
}

\begin{abstract}

The LHCb Experiment is preparing a detector upgrade fully exploit the
flavour physics potential of the LHC. The whole detector will be read
out at the full collision rate and the online event selection will be
performed by a software trigger. This will increase the event yields
by a facto $10$ for muonic and a factor $20$ for hadronic final
states. Research towards the upgrade has started with the target to
install the detector in 2018.

\end{abstract}

\maketitle


\section{Introduction}
\label{sec:intro}

The LHCb Experiments~\cite{bib:LHCb} is one of the four main
experiments at the LHC. LHCb is designed for precision measurements of
CP violation and rare decays of $b$ and $c$ flavoured hadrons, but has
broadened its programme to a wide range particle physics topics. Its
forward spectrometer geometry enables world leading measurements both
within and outwith the original core programme.

LHCb has successfully collected collision data since 2010 and has to
date collected more than $2~\rm{fb^{-1}}$ of integrated
luminosity. More than 92 \% of the delivered luminosity is recorded
with the detector fully operational and 99 \% of the recorded data
passes the quality criteria for use in offline analyses.

The experiment covers the pseudo-rapidity region of $2<\rm{\eta}<5$
and is fully instrumented in the whole acceptance range. LHCb profits
from the very large heavy flavour production cross sections at
LHC. Moreover, the majority of the $b$ and $c$ flavoured hadrons are
produced in the very forward region. Even though LHCb only covers 2 \%
of the stereo angle, 27 \% of the $b\bar{b}$ pairs are produced within
its acceptance. The production cross sections within the detector
acceptance are $\sigma_{b\bar{b}}=75.3~\rm{\mu b}$~\cite{bib:bbbarxs}
and $\sigma_{c\bar{c}}=1.23~\rm{mb}$~\cite{bib:ccbarxs} for proton
collisions at $7~\rm{TeV}$.

Hence LHCb has access to a sample of heavy flavour decays of
unprecedented size. As an illustration, approximately $8\cdot10^{10}$
$b\bar{b}$ pairs were produced in the LHCb acceptance in the 2011 run
($1~\rm{fb^{-1}}$ integrated luminosity), which is more than 50 times
larger than the total number of $b\bar{b}$ pairs produced in the
B-factories (Belle and BaBar) combined. This comes at the price of a
more challenging experimental environment. The total cross
section~\cite{bib:totem} is several orders of magnitude larger than
the $b\bar{b}$ cross section, hence the signal candidates have to be
selected among a large number of background events. Moreover, the
track multiplicity for events containing the signal candidates is
larger than in the B-Factories. Hence the event selection is crucial
for the performance of the experiment, in particular the online
selection performed by the trigger system.
 
LHCb has shown that it performs very well in these challenging
experimental conditions and that high precision measurements are
possible at at a hadron collider. This has already resulted in a large
number of publications, many of them representing world best
measurements or first observations. 

LHCb is currently operating at a luminosity of
$4\cdot10^{32}~\rm{cm^{-2}s^{-1}}$, which is twice the design value
but still significantly below what LHC could provide. A scheme of
luminosity levelling has been devised where the beams are slightly
displaced from head-on collisions and gradually re-centred during the
fill as the beam intensity is reduced. This allows the experiment to
collect data at a constant luminosity throughout the fill. Hence LHCb
is not limited by the luminosity that the LHC can deliver, but rather
by the luminosity that could be turned in to high precision
measurements.

The aim of the LHCb Upgrade~\cite{bib:fwtdr} is to build an experiment
that can operate at a higher luminosity while maintaining or
increasing the efficiency of selecting signal candidates. The details
of the upgraded experiment are outlined in this paper.

\section{The Current and Upgraded Trigger}
\label{sec:physics}

The current online event selection (trigger) consists of two levels. A
first level (L0) is is selecting events with large energy depositions
and high momentum muon tracks using information from the calorimeters
and muon system only. It is implemented in hardware and is reducing
the rate from the up to $30~\rm{MHz}$ collision rate to approximately
$900~\rm{kHz}$ event rate. The total latency for the L0 trigger
decision is $4~\rm{\mu s}$. The full detector information is read out
for the selected events and sent to the High Level Trigger (HLT)
system which is software based and implemented in a CPU farm. The
events are fully reconstructed and those containing candidates
matching inclusive and exclusive signal criteria are written to tape
for offline analysis. The output event rate of the HLT is approximately
$4.5\rm{kHz}$.

The goal for the LHCb upgrade is to operate the detector at a
luminosity of $1-2\cdot10^{33}~\rm{cm^{-2}s^{-1}}$ to collect up to
$50~\rm{fb^{-1}}$ during 5 years of data taking. Unfortunately the
signal yields would saturate if the luminosity were increased to these
levels with the current experiment. This is due to the limited
information and short time available for the L0 trigger decision. The
effect is particularly prominent for signal with hadronic final
states, as the muonic triggers scale better with increased luminosity.

Hence the strategy of the LHCb upgrade is to remove the L0 trigger and
move to a fully software based trigger with the full event information
available for all events. The aim is to reduce the $30~\rm{MHz}$
collision rate to $20~\rm{kHz}$ event rate written to tape. The option
of a Low Level Trigger (LLT) is currently being investigated, running
a fast event selection using non-CPU based processing based on for
instance digital signal processors or graphical processing units. This
would allow a gradual increase of the input rate to the HLT as the
available computing resources increase, finally reaching the full
$30~\rm{MHz}$ rate. The goal is to increase the yield of muonic
channels with a factor $10$ and for hadronic channels with a factor
$20$ compared to the current experiment.

\section{Detector Upgrade}
\label{sec:detector}

The change of trigger strategy will require an upgrade of most of the
on- and off-detector electronics to provide read-out at the full
collision rate for all systems. In addition, the increased luminosity
will lead to larger track densities and higher radiation dose on the
detector components. Hence a large fraction of the detector hardware
needs to be upgraded.

One of the principal challenges is to read out and process the copious
amount of data produced. Significant data reduction in the front-end
electronics is necessary to reduce the bandwidth required in the data
links off detector. This will be done with custom algorithms tailored
to each detector system. Nevertheless a massive network of optical
links will be necessary and the off detector electronics will have to
receive, re-package and re-distribute tenths of $\rm{Tb/s}$ of data to
the trigger system.

Research is ongoing to study the detector layout and to develop
components for the upgrade. Several different design options are being
evaluated, with many key choices to be maid in the coming years. A few
details on the upgrades of the different systems are detailed below.

The Vertex Locator is investigating both a fine-pitched silicon strip
detector and a hybrid pixel detector, with a reduced inner radius to
minimise the extrapolation distance. Common challenges are to
implement the on-chip zero-suppression in the front-end ASIC and to
handle the large radiation dose, up to $5\cdot10^{15}~1~\rm{MeV}$
neutron equivalent fluence.

The tracking system will use a combination of silicon strip sensors
and scintillating fibres while keeping the current straw tube tracking
for the large radii. Common developments for the front-end electronics
between the tracking systems and the Vertex Locator are being
investigated.

The photon detectors for the RICH detectors have to be replaced to
provide read-out at the full collision rate. A possible extension to
the baseline design is a combined time-of-flight and RICH detector
using photon detectors with very high timing resolution. This would
replace one of the RICH detectors for low momentum particle
identification.

The impact of the upgrade is smallest on  the calorimeter and muon
systems as they are already part of the L0 trigger. However, the
front-end electronics need to be replaced to cope with the higher data
rates and the first muon station will be removed as it will be
saturated by the high track multiplicity. 

The target is to install and commission the upgraded detector in the
second long LHC shut down, scheduled for 2018-2019. 





\bibliographystyle{aipproc}   

\bibliography{287_Eklund}

\IfFileExists{\jobname.bbl}{}
 {\typeout{}
  \typeout{******************************************}
  \typeout{** Please run "bibtex \jobname" to optain}
  \typeout{** the bibliography and then re-run LaTeX}
  \typeout{** twice to fix the references!}
  \typeout{******************************************}
  \typeout{}
 }

\end{document}